\documentclass[conference]{IEEEtran}
\IEEEoverridecommandlockouts
\usepackage{amsmath,amssymb,amsfonts}
\usepackage{algorithmic}
\usepackage{graphicx}
\usepackage{textcomp}
\usepackage{xcolor}
\usepackage{hyperref}
\usepackage{siunitx}
\def\BibTeX{{\rm B\kern-.05em{\sc i\kern-.025em b}\kern-.08em
 T\kern-.1667em\lower.7ex\hbox{E}\kern-.125emX}}
 
\usepackage[noadjust]{cite}

\newcommand{\TOMOGAN}{TomoGAN}
\newcommand{\jetson}{\texttt{TX2}}
\newcommand{\coraldb}{\texttt{Dev Board}}
\newcommand{\coralacc}{\texttt{Accelerator}}
\newcommand{\pqt}{\texttt{Post-Quantization}}
\newcommand{\qtaw}{\texttt{Quantization-Aware}}
\newcommand{\edgetpu}{\texttt{Edge-TPU}}
\newcommand{\mobilecomp}{\texttt{Mobile-Compatible}}
\newcommand{\edgetpucomp}{\texttt{Edge-TPU-Compatible}}

\newif\iffinal

\iffinal
 \newcommand\ian[1]{}
 \newcommand\zliu[1]{}
 \newcommand\vibhatha[1]{}
 \newcommand\raj[1]{}
 \newcommand\relocate[1]{}
 \newcommand\er[1]{}
 \newcommand\inpr[1]{}
 \newcommand\deprc[1]{}
 \newcommand\rrw[1]{}
 \newcommand\rpt[1]{}
 \newcommand\rcmt[1]{}
\else
 \newcommand\ian[1]{{\color{red}[ Ian: #1 ]}}
 \newcommand{\zliu}[1]{{\textcolor{blue}{ Zhengchun: #1 }}}
 \newcommand{\vibhatha}[1]{{\textcolor{orange}{ Vibhatha: #1 }}}
 \newcommand\raj[1]{{\color{green}[ Raj: #1 ]}}
 \newcommand\relocate[1]{{\color{purple}[ [RELOCATE]: #1 ]}}
 \newcommand\er[1]{{\color{magenta}[ \textbf{[ERROR]}: #1 ]}}
 \newcommand\inpr[1]{{\color{teal}[ \textbf{[IN-PROGRESS]}: #1 ]}}
 \newcommand\deprc[1]{{\color{brown}[ \textbf{[DEPRECATED]}: #1 ]}}
 \newcommand\rrw[1]{{\color{cyan}[ \textbf{[READY-TO-REVIEW]}: #1 ]}}
 \newcommand\rpt[1]{{\color{olive}[ \textbf{[REPITITIVE-MATERIAL]}: #1 ]}}
 \newcommand\rcmt[1]{{\color{darkgray}[ \textbf{[COMMENT-REQUIRED]}: #1 ]}}
\fi

\begin{document}
\title{Scientific Image Restoration Anywhere}
\author{
\IEEEauthorblockN{Vibhatha Abeykoon\IEEEauthorrefmark{1},
Zhengchun Liu\IEEEauthorrefmark{2},
Rajkumar Kettimuthu\IEEEauthorrefmark{2},
Geoffrey Fox\IEEEauthorrefmark{1} and
Ian Foster\IEEEauthorrefmark{2}\IEEEauthorrefmark{3}}
\IEEEauthorblockA{\IEEEauthorrefmark{1}School of Informatics, Computing and Engineering,
IN 47405, USA\\
\{vlabeyko, gcf\}@iu.edu}
\IEEEauthorblockA{\IEEEauthorrefmark{2}Data Science and Learning Division,\
Argonne National Laboratory, Lemont, IL 60439, USA\\
\{zhengchung.liu, kettimut, foster\}@anl.gov}
\IEEEauthorblockA{\IEEEauthorrefmark{3}Department of Computer Science, 
University of Chicago, Chicago, IL 60637, USA}
}

\maketitle
\begin{abstract}
The use of deep learning models within scientific experimental facilities frequently requires low-latency inference, so that, for example, quality control operations can be performed while data are being collected. Edge computing devices can be useful in this context, as their low cost and compact form factor permit them to be co-located with the experimental apparatus.
Can such devices, with their limited resources, can perform neural network feed-forward computations efficiently and effectively? We explore this question by evaluating the performance and accuracy of a scientific image restoration model, for which both model input and output are images, on edge computing devices. Specifically, we evaluate deployments of TomoGAN, an image-denoising model based on generative adversarial networks developed for low-dose x-ray imaging, on the Google Edge TPU and NVIDIA Jetson. We adapt TomoGAN for edge execution, evaluate model inference performance, and propose methods to address the accuracy drop caused by model quantization. We show that these edge computing devices can deliver accuracy comparable to that of a full-fledged CPU or GPU 
model, at speeds that are more than adequate for use in the intended deployments,
denoising a 1024$\times$1024 image in less than a second. 
Our experiments also show that the Edge TPU models can provide 3$\times$ faster inference response than a CPU-based model and 1.5$\times$ faster than an edge GPU-based model.
This combination of high speed and low cost permits image restoration anywhere.
\end{abstract}


\section{Introduction}
\noindent
Deep neural networks show considerable promise for the rapid analysis of data collected at scientific experiments,
enabling tasks such as anomaly detection~\cite{chandola2009anomaly}, image enhancement~\cite{yang2018low-enhance}, and image reconstruction~\cite{schlemper2017deep} to be performed in a fraction of the time
required by conventional methods.
However, the substantial computational costs of deep models are an obstacle to their widespread adoption.
The graphical processing unit (GPU) devices that are typically used to run these models are too expensive to be
dedicated to individual instruments, 
while dispatching analysis tasks to shared data centers can require substantial data movement and incur large round trip latencies. 

Specialized ``edge'' inference devices~\cite{xu2018scaling} such as the NVIDIA Jetson Tx2 GPU (henceforth \jetson{}) \cite{nvidia-jetson-Tx2} 
and Google Edge TPU~\cite{google-coral}
are potential solutions to this problem. 
(The Edge TPU is distributed by Coral Inc.\ in two configurations: \coralacc{}, which relies on a host machine, such as a PC or single-board Raspberry Pi, and \coraldb{}, which comes with a 64-bit ARM system as host.)
These edge devices use techniques such as reduced precision arithmetic to enable rapid execution of deep models
with a low price point (and low power consumption and compact form factor) that makes it feasible to embed them within scientific instruments. 

The question remains, however, as to whether these edge inference devices can execute the deep models used in science with sufficient speed
and accuracy.
Models originally developed to run on GPUs that support 32-bit floating point arithmetic must be adapted to run on edge devices that may support only
lower-precision integer arithmetic, a process that typically employs a technique called model quantization \cite{park2018value,hou2018loss,krishnamoorthi2018quantizing}. 
Google and NVIDIA have developed implementations of such schemes, allowing inference with integer-only arithmetic on integer-only hardware \cite{google-quantization, migacz2017nvidia-tensorrt, tflite-google}. 
They report benchmark results showing that edge devices can perform inference as rapidly as a powerful PC, at much lower cost \cite{coral-google-benchmark,nvidia-benchmarks}.
However, questions remain when it comes to using such devices in scientific settings.
The resulting models will be more compact, but will they be sufficiently accurate? 
And will the edge device run the models rapidly enough to meet scientific goals?

We explore these questions here by studying how a specific scientific deep learning model, \TOMOGAN{}\footnote{Code available at \url{git@github.com:ramsesproject/TomoGAN.git}}~\cite{tomogan-anl,tomogan-stream},
can be adapted for edge deployment.
\TOMOGAN{} uses generative adversarial network (GAN) methods~\cite{gan} to enhance the quality of low-dose X-ray images via a denoising process. 
Although diverse object detection and classification applications have been implemented on edge devices, 
image restoration with a complex image generative model has not previously been attempted on them. 
We adapt TomoGAN to run on the Google Edge TPU (both \coralacc{} and \coraldb{}) and \jetson{}, and compare the accuracy and computational performance of the resulting models with those of other implementations.  
We also describe how to mitigate accuracy loss in quantized models by applying a lightweight ``fine-tuning'' convolutional neural network 
to the results of the quantized TomoGAN. 

The rest of this paper is as follows. 
In \S\ref{s:methodology}, we describe how we adapt a pre-trained deep learning model, \TOMOGAN{}, for the Edge TPU.
Next in \S\ref{s:experiments}, we present experiments used to evaluate edge computing performance and model accuracy,
both with and without the fine-tuning component. 
In \S\ref{s:related-work} we review related work, and in \S\ref{s:conclusion} we summarize our results and outline directions for future work.  
\section{Methodology}\label{s:methodology}
\noindent
We next describe how we adapt the \TOMOGAN{} model for the \edgetpu{} (i.e., \coralacc{} and \coraldb{}). 
Specifically, we describe the steps taken to improve the accuracy of the enhanced images, the datasets used, and our performance and accuracy evaluations. 

\subsection{Quantization}
\noindent
We consider two approaches to quantizing 
the \TOMOGAN{} model for the Edge TPU: \pqt{} and \qtaw{}. In both methods, the first step is to design a non-quantized model with the expected features unique to both \pqt{} and \qtaw{} models. 

\subsubsection{Post-Quantization-Based Inference Model}\label{ss:pqt}
\begin{figure}[htbp]
  \centering
  \includegraphics[width=\columnwidth]{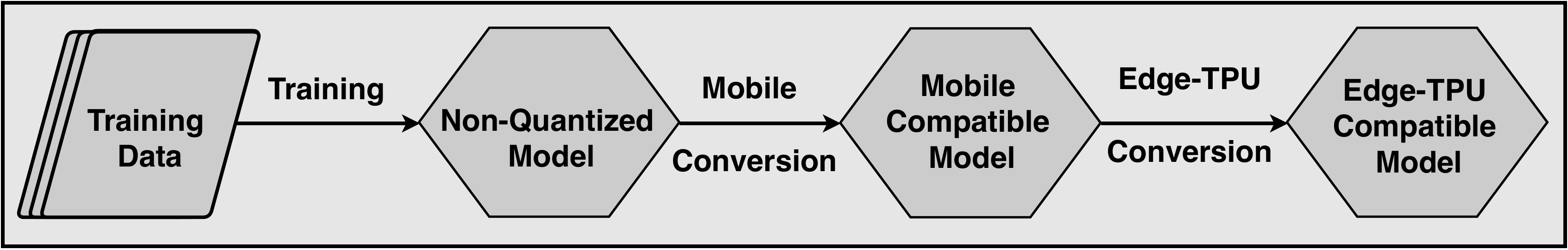}
  \caption{The \pqt{} model generation workflow involves a training step, conversion of the trained model to a mobile form, and then
  conversion of the mobile form to an Edge TPU model.}
  \label{fig:tomogan-post-quantization-workflow}
\end{figure}

\noindent
The steps followed to generate the \pqt{}-based inference model are shown in 
 \autoref{fig:tomogan-post-quantization-workflow}.

We first train a \pqt{}-based model, which differs from the standard \TOMOGAN{} model only in the input tensor shape,
which is 64$\times$64$\times$3 rather than 1024$\times$1024$\times$3. 
The partitioning of each 1024$\times$1024 input image into multiple 64$\times$64 subimages is needed because of limitations on the output size
permitted by the \edgetpucomp{}.
See \S\ref{ss:datasets} for details on training data.
The average training time for 40,000 iterations was around 24 hours on a single NVIDIA V100 GPU.

The resulting trained \pqt{} model is then converted to a \mobilecomp{} model (see \S\ref{sss:mobile-compatible}),
which is used in term to generate an \edgetpucomp{} model (see \S\ref{sss:edgetpu-comp-model-gen}).

\subsubsection{Quantization-Aware Based Inference Model}\label{sss:qtaw-inference-model}
\begin{figure}[htbp]
  \centering
  \includegraphics[width=\columnwidth]{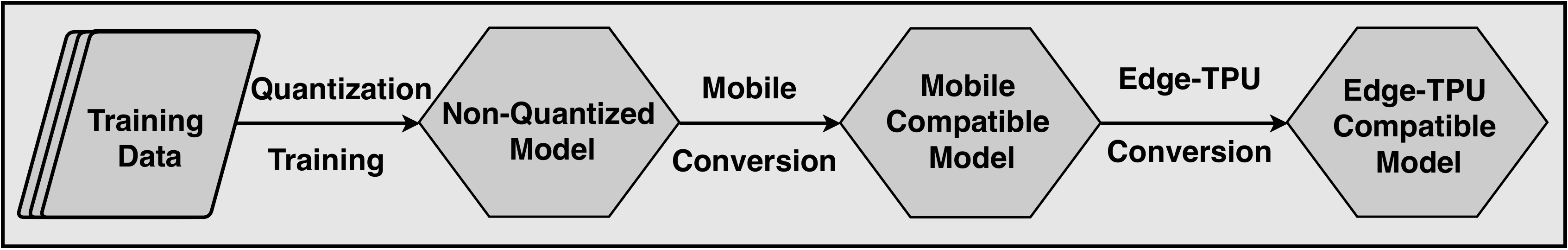}
  \caption{The \qtaw{} model generation workflow differs from the  \pqt{} workflow only in its training step.}
  \label{fig:tomogan-quantization-aware-workflow}
\end{figure}

This second approach (\S\ref{fig:tomogan-quantization-aware-workflow}) differs from the first only in the method used to generate the trained model.
In order to attenuate the accuracy loss that may result from the quantization of trained weights in the inference stage,
a more complex model is trained that induces fake quantization layers to simulate the effect of quantization. 

The major drawback of this methodology is that the introduction of fake quantization layers leads to a much longer training time, extending to dozens of days.  
We thus conclude that this method is not feasible for larger models like \TOMOGAN{},
and 
adopt the \pqt{} approach for our \TOMOGAN{}-based image restoration system. 

\subsection{Model Generation}
\noindent
There are specific model generation schemes to provide mobile-computing and edge-computing friendly models. 

\subsubsection{Mobile-Compatible Model Generation}\label{sss:mobile-compatible}
A mobile-compatible model accepts quantized \texttt{unsigned int8} inputs and generates quantized \texttt{unsigned int8} outputs.
A quantized \emph{int8\_value} representation is related to the corresponding \emph{real\_value} as follows
\begin{equation*}\label{eq:qauntization}
  real\_value = (int8\_value - zero\_point) * scale,
\end{equation*}
where \emph{zero-point} and \emph{scale} are parameters.
Prior to mobile-compatible model generation,
we process a representative dataset to estimate the value range of the data that are to be quantized, and choose
appropriate values for these parameters.

\subsubsection{Edge-TPU Compatible Model Generation}\label{sss:edgetpu-comp-model-gen}
In order to exploit the \coralacc{} and \coraldb{}, the quantized model must be converted into an \edgetpucomp{} model by compiling it with the Edge TPU runtime. \edgetpucomp{} model generation is done by using a compiler deployed with \edgetpu{} firmware libraries. This compiler enables a conversion of a \mobilecomp{} model into an \edgetpucomp{} model. 

\subsection{Inference Workflow} \label{s:inf-wf}

We describe in turn the inference workflows used when running on a CPU, \edgetpu{}, and Edge GPU.
The CPU related experiments are carried out with the trained model with no quantization and the \edgetpu{} and Edge-GPU experiments are carried out with the trained model with quantization.

\subsubsection{CPU Inference}\label{sss:cpu-inference-workflow}

\begin{figure}[htbb]
  \centering
  \includegraphics[width=0.8\columnwidth]{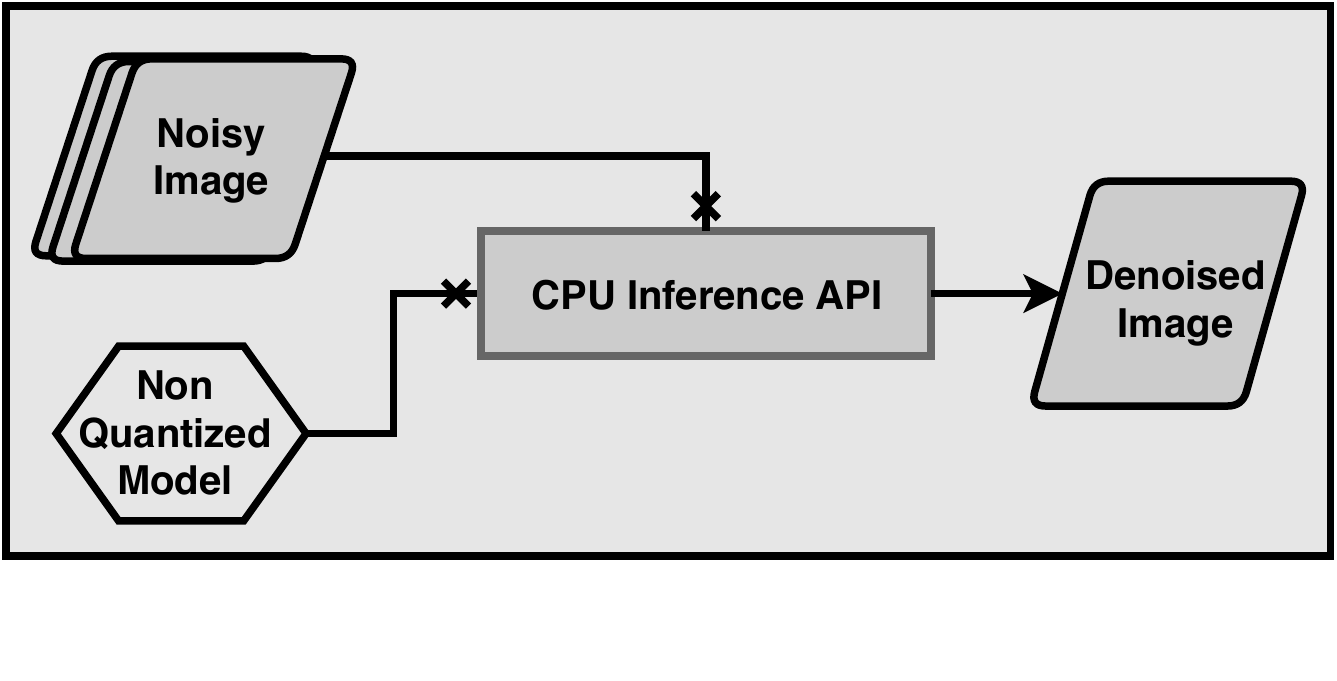}
\vspace{-5ex}
  \caption{The CPU inference workflow applies the non-quantized TomoGAN direclty, with no pre- or post-processing required.}
  \label{fig:cpu-inference-workflow}
\end{figure}

The CPU inference workflow, shown in
\autoref{fig:cpu-inference-workflow}, uses the non-quantized model. 
We feed the required inputs, non-quantized model, and noisy image of size 1$\times$1024$\times$1024$\times$3
 to the CPU-based inference API, which 
returns a de-noised image with dimension 1024$\times$1024. 

\subsubsection{Edge TPU Inference}\label{sss:tpu-inference-workflow}

\begin{figure}[htbb]
  \centering
  \includegraphics[width=\columnwidth]{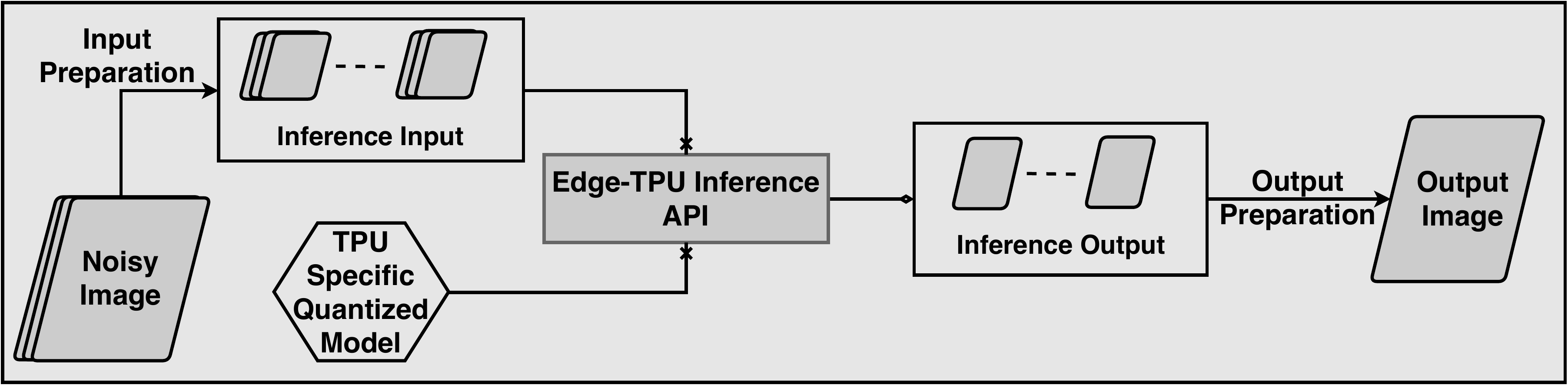}
  \caption{The Edge TPU inference workflow partitions each input image into 256 subimages,
  processes the subimages, and stitches the 
  corresponding output subimages together to create the output image.}
  \label{fig:edge-tpu-inference-workflow}
\end{figure}

\noindent
The Edge TPU inference workflow, shown in 
\autoref{fig:edge-tpu-inference-workflow},
applies the \pqt{} (\ref{ss:pqt}) or \qtaw{} (\ref{sss:qtaw-inference-model}) models to preprocessed images.
We use customized versions of the \coralacc{} and \coralacc{} \texttt{BasicEngine} inference API and \texttt{TensorflowLite} API \cite{tflite-google} for this purpose.
Each input image has shape 1$\times$1024$\times$1024$\times$3, where the dimension 3 results from grouping with each image two adjacent images,
as used in \TOMOGAN{} to improve output quality.
Each image is partitioned into 256 \emph{subimages} of shape 1$\times$64$\times$64$\times$3, due to \edgetpucomp{} restrictions;
after inference, processed subimages are buffered in memory, and once all have been processed, are stitched back together to form the de-noised output image. 

\subsubsection{Edge-GPU Based Inference Workflow}\label{sss:edge-gpu-inferecne-workflow}

\begin{figure}[htbb]
  \centering
  \includegraphics[width=\columnwidth]{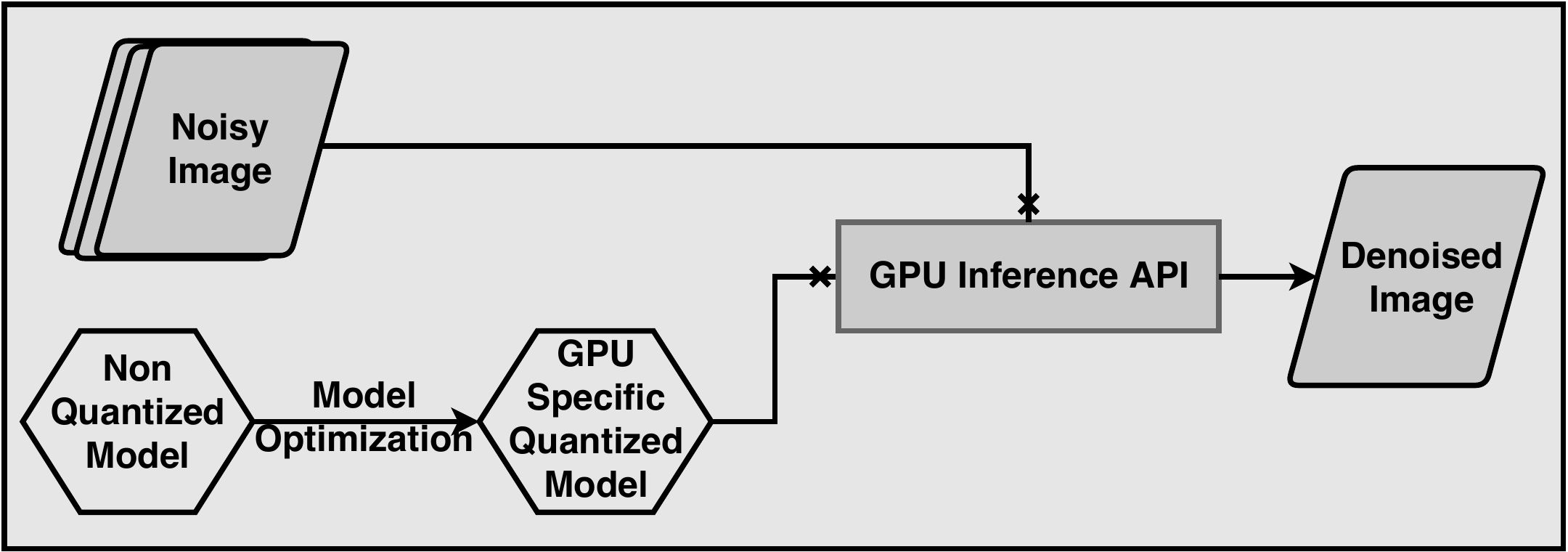}
  \caption{The Edge GPU inference workflow differs from the CPU inference workflow in its application of model optimization to
  the non-quantized model.}
  \label{fig:edge-gpu-inference-workflow}
\end{figure}

The Edge GPU inference workflow, shown in \autoref{fig:edge-gpu-inference-workflow}, is similar to the CPU inference workflow, except for the part of using a GPU specific quantized model.
Each input image, with shape 1$\times$1024$\times$1024$\times$3,
is passed to the Edge GPU-specific quantized model (produced with TensorRT \cite{tensorrt} from the non-quantized model), 
which produces a de-noised image with shape 1024$\times$1024 as output.  

\subsection{Fine-Tuning Workflow}\label{s:fine-tune-workflow}

\begin{figure}[htbp]
  \centering
  \includegraphics[width=\columnwidth]{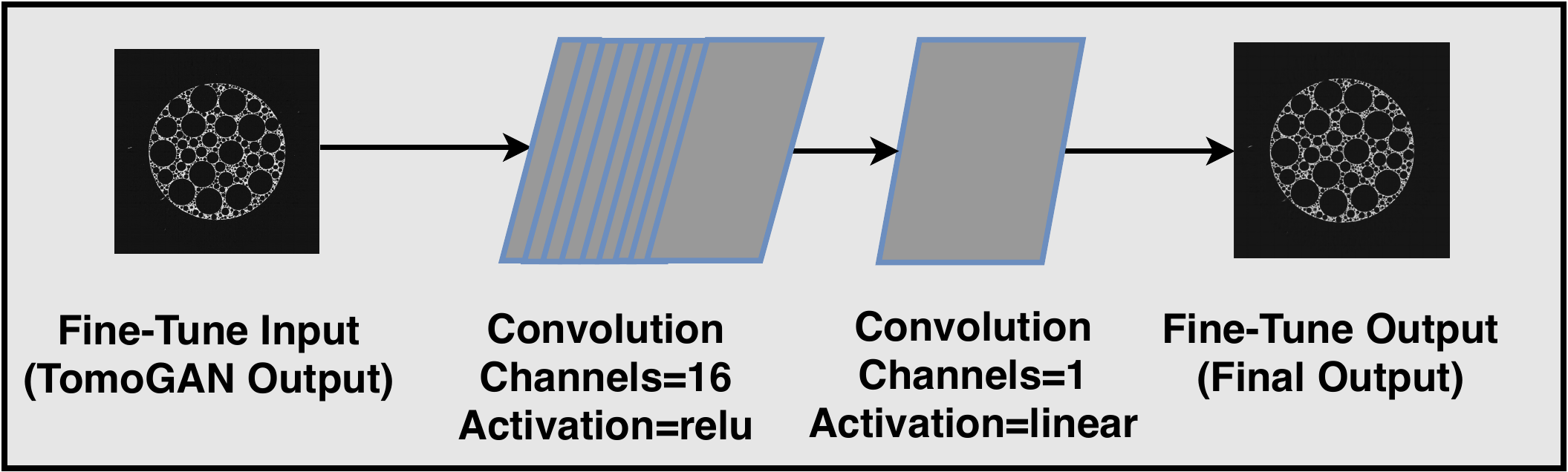}
  \caption{The Fine-Tune network is applied to the output from the quantized TomoGAN to improve image quality.}
  \label{fig:fine-tune-network}
\end{figure}

\noindent
With \pqt{}-enabled inference, some accuracy may be lost due to model quantization.
We observed this effect in our preliminary results:
the non-quantized model produced better output than the  \pqt{} Edge-TPU quantized model. 
To improve image quality in this case, we designed a shallow convolutional neural network (referred to as the \emph{Fine-Tune network} in the rest of the paper) to be applied to the output of the quantized \TOMOGAN{}: see \autoref{fig:fine-tune-network}.
We use output from the \edgetpucomp{} model (see \S\ref{sss:edge-gpu-inferecne-workflow} and \S\ref{ss:datasets}) to train this network.
The target labels are the corresponding target images for each inferred image from the mentioned portion of the training dataset. 
At the inference stage, we applied the \edgetpu{} inference workflow (see \S\ref{sss:edge-gpu-inferecne-workflow}) and used its output as input to the Fine-Tune model.
We shall see in \S\ref{ss:image-quality} that this Fine-Tune network 
improves image quality to match that of the images generated from the CPU inference workflow 

\section{Experiments}\label{s:experiments}
\noindent
In order to evaluate both computing throughput and model inference performance, we conducted a set of experiments on quantized inference and enhanced the output from direct inference with a shallow Fine-Tune network. 
We compared the performance and evaluated the image quality on CPU, GPU, and TPU devices individually. 

\subsection{Datasets}\label{ss:datasets}
\noindent
We used two datasets for our experiments. Each dataset comprises 1024 pairs of 1024$\times$1024 images, 
each pair being a noisy image and a corresponding ground-truth image, as described in Liu et al.~\cite{tomogan-anl}.
Ground truth images are obtained from normal-dose X-ray imaging and noisy images from low-dose X-ray imaging of the same sample. 
We used one dataset for training and the other for testing. 

\subsection{Performance Evaluation}
\noindent
We evaluated both inference performance~(i.e., throughput) on different hardware platforms and the quality of the resulting images. 
For inference performance, we studied a laptop CPU (\S\ref{sss:cpu-hardware}), the \coralacc{} and \coraldb{} Edge TPU (\S\ref{sss:tpu-hardware}), and the \jetson{} Edge GPU (\S\ref{sss:gpu-hardware}),
applying for each the workflow of \S\ref{s:inf-wf} to a series of images and 
calculating the average inference latency.

\subsubsection{CPU Inference Performance Evaluation}\label{sss:cpu-hardware}
\noindent
Standard CPU-based experiments were conducted by using the non-quantized model with a personal computer comprising an Intel Core i7-6700HQ CPU@2.60GHz with 32GB RAM. 
The supported operating system was Ubuntu 16.04 LTS distribution. 
The non-quantized model takes an average inference time of 1.537 seconds per image:
see \texttt{i7@2.6GHz} in \autoref{fig:inference-latency}. 

\subsubsection{TPU Inference Performance Evaluation}\label{sss:tpu-hardware}
\noindent
We evaluated Edge TPU performance on two platforms with different configurations:
the \coralacc{} with an Edge TPU coprocessor connected to the host machine (a laptop with Intel i7 CPU) via a USB 3.0 Type-C (data and power) interface,
and the \coraldb{} with Edge TPU coprocessor and a 64-bit ARM CPU as host. 
Columns \texttt{Accelerator} and \texttt{Dev Board} of
\autoref{tb:time-breakdown-edge} provide timing breakdowns for these two devices. 
The first component is the time to run the quantized TomoGAN model: 
0.435 and 0.512 seconds per image for \coralacc{} and \coraldb{}, respectively. 
The second component, ``Stitching,'' is due to an input image size limit of 64$\times$64  imposed by the 
Edge TPU hardware and compiler that we used in this work. 
Processing a single 1024$\times$1024 image thus requires processing 256 individual 64$\times$64 images,
which must then be stitched together to form the complete output image.
This stitching operation
takes an average of 0.12 and 0.049 seconds per image on the \coraldb{} and \coralacc{}, respectively.

The third component, ``Fine-Tune,'' is the quantized fine-tune network used to improve image quality to match that of the non-quantized model, as discussed in \S\ref{ss:image-quality}; this takes an average of 0.070 and 0.166 seconds per image on \coralacc{} and \coraldb{}, respectively. 
We note that model compilation limitations associated with the current Edge TPU hardware and software 
require us to run the quantized \TOMOGAN{} and Fine-Tune networks separately,
which adds extra latency for data movement between host memory and Edge TPU.
We expect to avoid this extra cost in the future by chaining \TOMOGAN{} and Fine-Tune to execute as one model. 
(While the quantized \TOMOGAN{} requires 301 billion operations to process a 1024$\times$1024 image,
Fine-Tune takes only  621 million: a negligible 0.2\% of \TOMOGAN{}.)

\begin{table}[htpb]
\caption{Performance Breakdown on Inference in Edge Devices. TOPS refers trillion operations  per second and TFLOPS denotes trillion floating point operations per second. }
\center
\begin{tabular}{l|c|c|c}
\noalign{\hrule height 2pt}
          & \multicolumn{1}{l|}{\textbf{Accelerator}} & \multicolumn{1}{l|}{\textbf{Dev Board}} & \multicolumn{1}{l}{\textbf{Jetson Tx2}} \\\noalign{\hrule height 2pt}
Quantized \TOMOGAN{} (s) & \textbf{0.435} & 0.512 & 0.880 \\ \hline
Stitching (s) & \textbf{0.049} & 0.120 & - \\ \hline
Fine-Tune (s) & \textbf{0.070} & 0.166 & - \\ \hline
Total (s) & \textbf{0.554} & 0.798 & 0.880 \\ \hline
Power Consumption (w) & 2 & 2 & 7.5\\ \hline
Peak Performance & 4 TOPS & 4 TOPS & 1.3 TFLOPS\\
\noalign{\hrule height 2pt}
\end{tabular}
\label{tb:time-breakdown-edge}
\end{table}

\subsubsection{Edge GPU Inference Performance Evaluation}\label{sss:gpu-hardware}
\begin{figure}[ht]
\center
\includegraphics[width=.8\columnwidth,trim=0 0.45in 0 0.1in,clip]{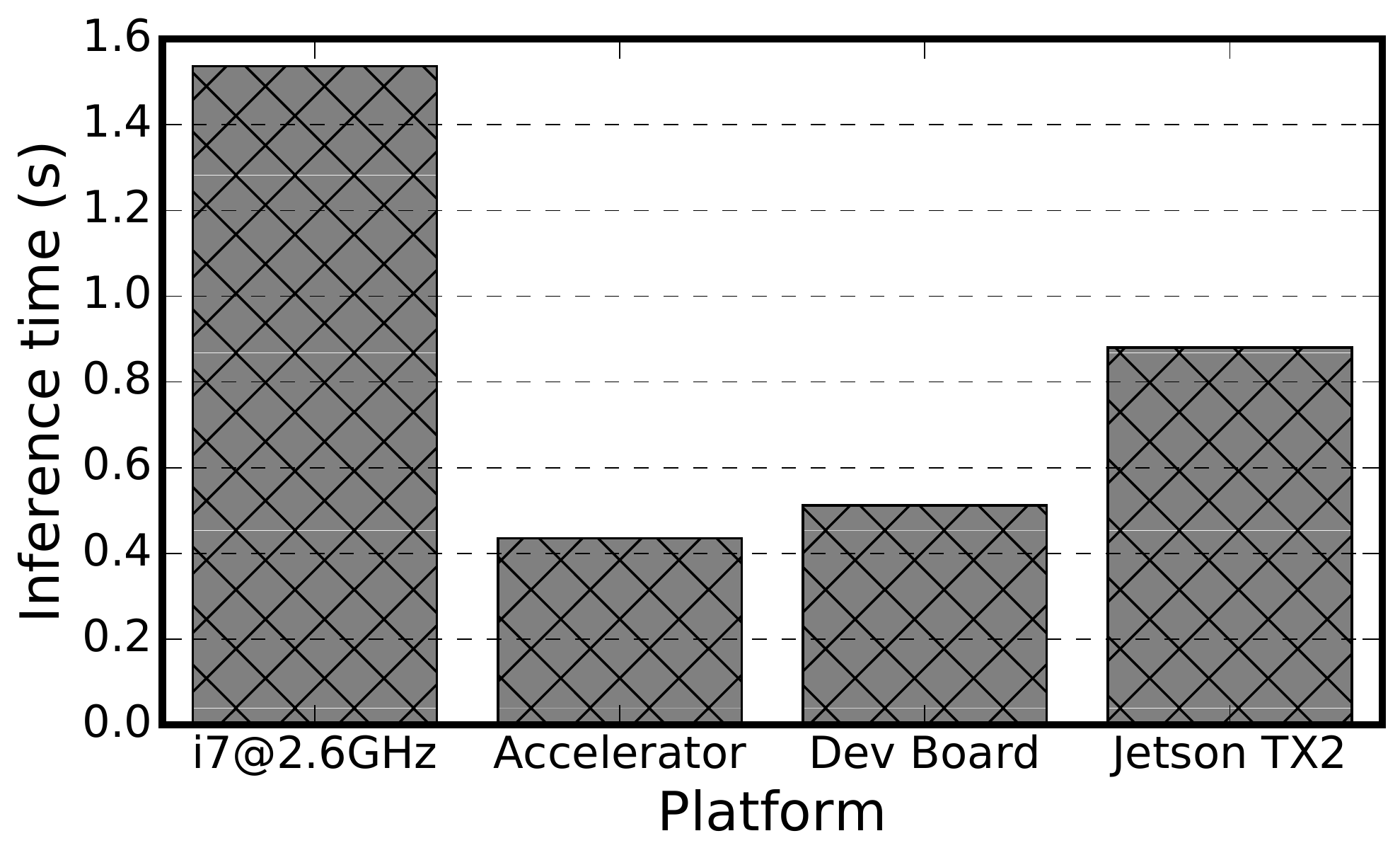}
\caption{Inference time for a 1024$\times$1024 pixel image on a CPU, the two Edge TPU platforms, and \jetson{}. 
}
\label{fig:inference-latency}
\end{figure}

The original \TOMOGAN{} can process a 1024$\times$1024 pixel image in just 44ms on a NVIDIA V100 GPU card.
As our focus here is on edge devices,
we evaluated TomoGAN performance on the \jetson{}, which has a GPU and is designed for edge computing. 
Column Jetson Tx2 in \autoref{tb:time-breakdown-edge} shows results.
We see an average inference time per image of 0.88 seconds for the \jetson{}.
We compare in \autoref{fig:inference-latency} this time with the quantized TomoGAN Edge TPU times (not including stitching and fine tuning).

We note that in constructing the model for the \jetson{}, 
we used NVIDIA's TensorRT toolkit~\cite{migacz2017nvidia-tensorrt, tensorrt} to optimize the operations of \TOMOGAN{}.
We also experimented with 16-bit floating point, 32-bit floating-point, and unsigned int8, and observed similar performance for each, which
we attribute to the lack of Tensor cores in the \jetson{}'s NVIDIA Pascal architecture for accelerating multi-precision operations.

\subsubsection{Performance Discussion}
We find that inference is significantly faster on the \edgetpu{} than on the CPU or \jetson{},
and faster on \jetson{} than on the CPU.
\coralacc{} is faster than \coraldb{},
because the former has a more powerful host (the laptop with i7) with better memory throughput than the latter (the 64-bit SoC ARM platform).
These performance differences may appear small, but we should remember that 
a single light source experiment can generate thousands of images, each larger than 1024$\times$1024, e.g., 2560$\times$2560, and thus any acceleration is valuable.

\subsection{Image Quality Evaluation}\label{ss:image-quality}
\begin{figure}[ht]
\centering
\includegraphics[width=0.8\columnwidth,trim=0 0.5in 0 0,clip]{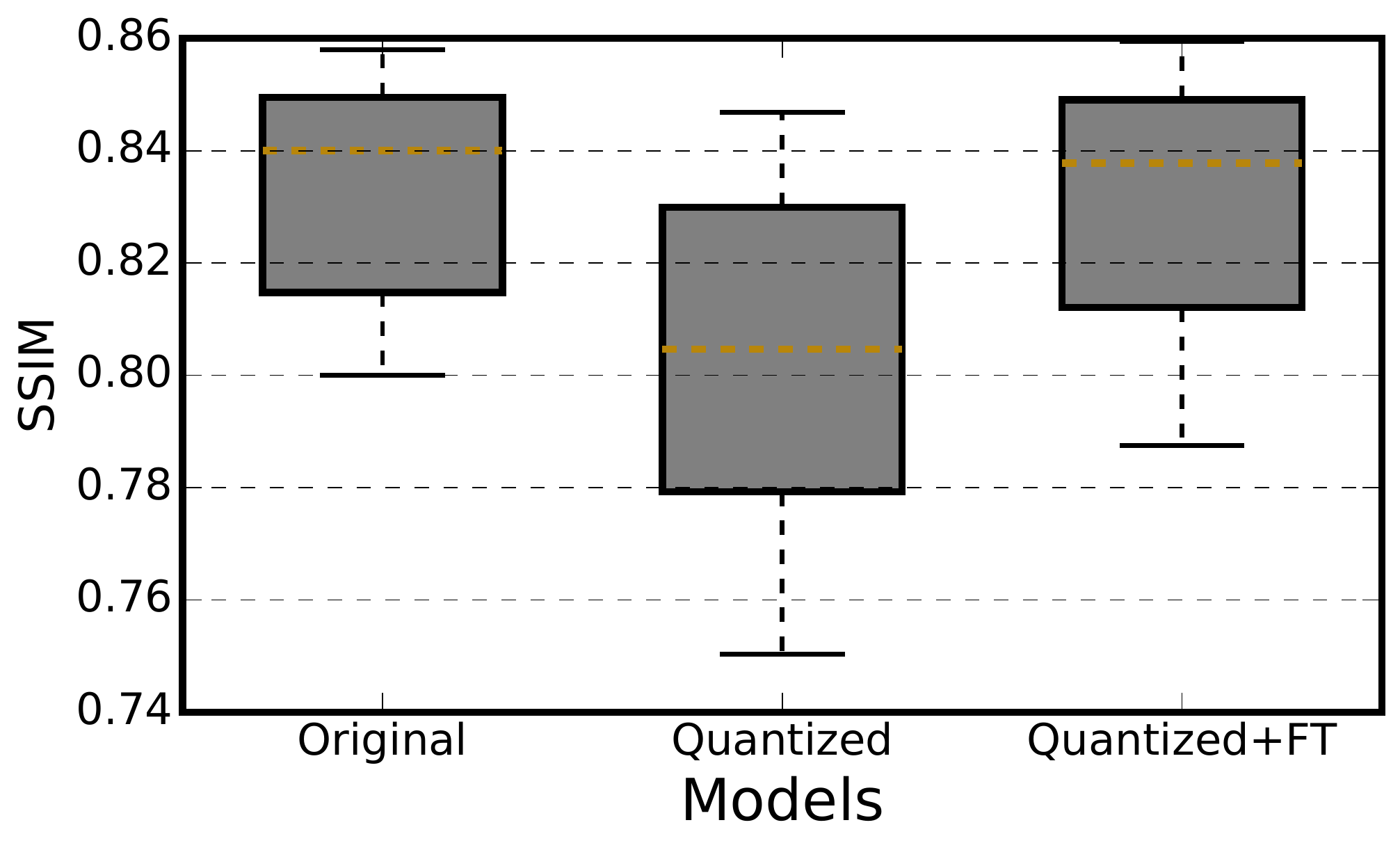}
\caption{SSIM image quality scores for 
 original TomoGAN,
  quantized TomoGAN, and
 quantized TomoGAN+Fine-Tune, each applied to our 1024 test images.
 (The boxes show 25th to 75th percentile values, and the dashed lines indicate the mean.)
 }
\label{fig:image-quality-ssim}
\end{figure}

\noindent
We used structural similarity index (SSIM)~\cite{wang2003ssim} to evaluate image quality.
We calculated this metric for images enhanced from the original TomoGAN, the quantized \TOMOGAN{}, and the quantized \TOMOGAN{} plus Fine-Tune network, with results shown in \autoref{fig:image-quality-ssim}.
We observe that SSIM for the quantized TomoGAN+Fine-Tune 
is comparable to that of the original (non-quantized) TomoGAN. 

\section{Related Work}\label{s:related-work}
\noindent
The opportunities and challenges of edge computing have received much attention~\cite{garcia2015edge,satyanarayanan2017emergence}.
Methods have been proposed for both co-locating computing resources with sensors,
and for offloading computing from mobile devices to nearby edge computers~\cite{chen2018thriftyedge,chen2018task,chen2015efficient}.


Increasingly, researchers want to run deep neural networks on edge devices~\cite{chen2019deep,zhou2019edge,lane2015can},
leading to the need to adapt computationally expensive deep networks for resource-constrained environments.
Quantization, as discussed above, is one such approach~\cite{google-quantization,hubara2017quantized,kim2016bitwise,hubara2016binarized}. 
Others include the use of neuromorphic hardware~\cite{krestinskaya2019neuromemristive}, 
specialized software~\cite{mobile-dnn},
the distribution of deep networks over
cloud, local computers, and edge computers~\cite{dnn-cloud-edge},
and mixed precision computations~\cite{DBLP:journals/corr/abs-1902-00460}.

Various deep networks have been developed or adapted for edge devices, including Mobilenet \cite{mobilenets}, VGG \cite{vggnet}, and Resnet~\cite{resnet}.
However, that work focuses on image classification and object detection. 
In contrast, we are concerned with image translation and image-to-image mapping to provide an enhanced image. 
Also, we are applying our image restoration model on edge devices, an approach that has not been discussed in the literature. 

Our use of a fine-tuning network to improve image quality is an important part of our solution, allowing us to avoid the excessive training time required for the quantization-aware model. We are not aware of prior work that has used such a fine-tuning network, although it is conceptually similar to the use of gradient boosting in ensemble learning~\cite{breiman1997arcing}.

\section{Conclusion}\label{s:conclusion}
\noindent
We have reported on the adaption for edge execution of 
TomoGAN, an image-denoising model based on generative adversarial networks developed for low-dose x-ray imaging.
We ported TomoGAN to the Google Coral Edge TPU devices (\coraldb{} and \coralacc{}) and NVIDIA Jetson \jetson{} Edge GPU.
Adapting TomoGAN for the Edge TPU requires quantization.
We mitigate the resulting loss in image quality, as measured via the SSIM image quality metric,
by applying a fine-tune step after inference, with negligible computing overhead. 
We find that 
\coraldb{} and \coralacc{} provide 3$\times$ faster inference than a CPU, and that
\coralacc{} is 1.5$\times$ faster than \jetson{}. 
We conclude that edge devices can provide fast response at low cost,
enabling \emph{scientific image restoration anywhere}.

The work reported here focused on image restoration.
However, before images can be enhanced with \TOMOGAN{}, they must be reconstructed
from the x-ray images, for example by using filtered back projection (FBP)~\cite{kak2002principles}.
FBP is not computationally intensive: processing the images considered here using the TomoPy implementation~\cite{gursoy2014tomopy} takes about 400ms per image on a laptop with an Intel i7 CPU.
Nevertheless, for a complete edge solution, we should also run FBP on the edge device.
We will tackle that task in future work.

\section*{Acknowledgment}
\noindent
This work was supported in part by the U.S. Department of Energy, Office of Science, Advanced Scientific Computing Research, under Contract DE-AC02-06CH11357.
This research was accomplished when V. Abeykoon is an intern at Argonne National Laboratory under the supervision of Z. Liu.
We thank the JLSE management team at the Argonne Leadership Computing Facility and the Google Coral team for their assistance. 

\bibliographystyle{IEEEtran}
\bibliography{IEEEexample}

\section*{License}
\noindent
The submitted manuscript has been created by UChicago Argonne, LLC, Operator of Argonne National Laboratory (``Argonne"). Argonne, a U.S. Department of Energy Office of Science laboratory, is operated under Contract No. DE-AC02-06CH11357. The U.S. Government retains for itself, and others acting on its behalf, a paid-up nonexclusive, irrevocable worldwide license in said article to reproduce, prepare derivative works, distribute copies to the public, and perform publicly and display publicly, by or on behalf of the Government. The Department of Energy will provide public access to these results of federally sponsored research in accordance with the DOE Public Access Plan. http://energy.gov/downloads/doe-public-access-plan.

\end{document}